\documentclass[useAMS,usenatbib]{mn2e}
\usepackage{epsfig}
\usepackage{amsmath}
\usepackage{graphicx}
\usepackage{epstopdf}
\usepackage{subfigure}

\newcommand{\be}{\begin{equation}}
\newcommand{\beq}{\begin{equation}}
\newcommand{\ba}{\begin{eqnarray}}
\newcommand{\ee}{\end{equation}}
\newcommand{\eeq}{\end{equation}}
\newcommand{\ea}{\end{eqnarray}}
\newcommand{\msun}{\mbox{$M_{\odot}$}}
\newcommand{\rsun}{\mbox{$d_{\odot}$}}
\newcommand{\vcirc}{\mbox{$V_{circ}$}}
\newcommand{\rvir}{\mbox{$r_{vir}$}}
\newcommand{\mvir}{\mbox{$M_{vir}$}}
\newcommand{\kms}{\mbox{km s$^{-1}$}}
\newcommand{\plm}{\mbox{$\pm$}}
\newcommand{\Mpc}{{\rm Mpc}}
\newcommand{\kpc}{{\rm kpc}}
\newcommand{\apj}{ApJ}

\newcommand{\mnras}{MNRAS}
\newcommand{\aj}{AJ}

\newcommand{\pasj}{PASJ}

\def\lsim{~\rlap{$<$}{\lower 1.0ex\hbox{$\sim$}}}

\def\gsim{~\rlap{$>$}{\lower 1.0ex\hbox{$\sim$}}}

\title[Orbit of the LMC]{Implications of Recent Measurements of the Milky
Way Rotation for the Orbit of the Large Magellanic Cloud} \author[Shattow
\& Loeb]{Genevieve Shattow and Abraham Loeb\\
Harvard-Smithsonian Center for Astrophysics, 60 Garden St, Cambridge,
MA 02138} \date{\today}

\begin{document}

\maketitle

\label{firstpage}
\begin{abstract}
We examine the implications of recent measurements of the Milky Way
(MW) rotation for the trajectory of the Large Magellanic Cloud (LMC).
The $\sim 14\pm 6\%$ increase in the MW circular velocity relative to
the IAU standard of $220~\kms$ changes the qualitative nature of the
inferred LMC orbit.  Instead of the LMC being gravitationally unbound,
as has been suggested based on a recent measurement of its proper
motion, we find that the past orbit of the LMC is naturally confined
within the virial boundary of the MW. The orbit is not as tightly
bound as in models derived before the LMC proper motion was measured.

\end{abstract}


\section{Introduction}
\label{sec:intro}

Recently, Kallivayalil et al. (2006; hereafter K06) measured the
proper motion of the the Large Magellanic Cloud (LMC), and pioneered a
first assessment of its 3D velocity vector.  Based on this
measurement, Besla et al. (2007; hereafter B07) concluded that the LMC
was unlikely to have passed near the Milky Way (MW) before and was
most likely formed outside of the boundaries of the Galaxy, in
contrast to traditional scenarios (see references in van der Marel et
al. 2002; hereafter vdM02).  This new conclusion is intriguing given
that there are no other examples of massive gas-rich galaxies like the
LMC within the much bigger volume that separates the MW from its
neighboring galaxy, Andromeda (M31).  Other recent studies have also
found the B07 results difficult to accept and propose alternate
strategies of binding the LMC and MW either through MOND gravity (Wu
et al. 2008) or by giving the LMC and its smaller counterpart, the
Small Magellanic Cloud, a common halo (Bekki 2008).

A couple of years after K06 published their findings, Piatek et
al. (2008; hereafter P08) confirmed independently their results to
within one standard deviation, although at the lower end of the
inferred range of values.  Also, within the past five years, the
circular velocity of the MW and the distance between the Sun
and the galactic center have been updated by Reid \& Brunthaler (2004;
hereafter RB04) and Gillessen et al. (2008; hereafter GGE08),
respectively. These increased the likely circular velocity of the MW
from the IAU standard of $V_{circ}=220$ \kms~to 251$\plm15$ \kms. A
value of 220 \kms~now corresponds to a reduction of the best-fit value
by two standard deviations (and equivalent to moving the Sun a total
of 1 kpc closer to the center of the Galaxy).  Uemura et al. (2000)
used parallax measurements from Hipparcos and SKYMAP to obtain a
similar value of $V_{circ}=255\plm8$ \kms.

The rotation speed of the MW affects the analysis of the past LMC
orbit in two ways. First, because the proper motion of the LMC is
measured relative to the solar system which orbits the Galaxy, it is
necessary to know the rotational velocity of the Sun in order to
transform to the Galactocentric frame (see, e.g. vdM02).  Second, the
depth of the gravitational potential well of the MW (involving its
estimated mass and scale radius) depends on the normalization of its
rotation curve.  B07 adopted the IAU standard in their analysis
instead of the modified values for the Milky Way's rotation. In this
{\it Letter}, we examine the implications of the change in the MW
parameters for the LMC orbit (also with the updated P08 value).  In
the particular geometry of the LMC orbit, both of the above-mentioned
effects make the LMC more gravitationally bound to the MW owing to an
increase in $V_{circ}$.  Despite the small fractional magnitude of the
correction in circular velocity ($\sim 14\pm 6\%$), we find that the
qualitative nature of the LMC orbit changes. Instead of the LMC being
possibly unbound as suggested by B07, the $\sim 10\%$ decrease in the LMC
velocity and the $\sim 50\%$ increase in the MW mass lead us
to conclude that the LMC's past trajectory was probably confined
within the virial radius of the MW.  The apogalacticon distance of the
orbit is comparable to the MW virial radius, as expected for a
satellite that had formed at the outer edge of the Galactic halo.

Traditional studies of the LMC's orbit around the MW (e.g., Murai \&
Fujimoto 1980; Lin \& Lynden-Bell 1982; Gardiner et al. 1994; Lin et
al. 1995; Gardiner \& Noguchi 1996; vdM02; Bekki \& Chiba 2005;
Mastropietro et al. 2005; Connors et al. 2006) considered the MW as an
isolated galaxy.  While this might have been acceptable for studies
where the LMC's orbit was confined well within the MW halo, the high
LMC velocity measured by K06 and P08 implies (B07) that the
apogalacticon could extend beyond the edge of the MW's halo -- where
the gravitational influence of M31 is non-negligible (see Table 1 in
B07).  Thus, we also include the tidal effect of M31 in our
calculations.

In \S \ref{sec:method} we describe our adopted model for the mass
distribution of the MW halo. We then calculate the past LMC orbit (\S
\ref{sec:const}) and the effect of M31 on it (\S \ref{sec:M31}). Finally,
we discuss the implications of our results in \S \ref{sec:disc}.

\section{Method}
\label{sec:method}

Following B07, we adopt a Navarro, Frenk, \& White (1996; hereafter
NFW) mass profile for the dark matter distribution in the MW halo, and
include dynamical friction (Chandrasekhar 1943, Hashimoto 2003) in
calculating the LMC motion through the MW halo.  We also add the
gravitational potential of M31 in the form of another NFW profile at a
present-day Galactocentric distance of 780 kpc (McConnachie et
al. 2005; Cox \& Loeb 2008 and references therein). The full
gravitational potential $\Phi_{tot}$ as a function of radius $r$ in
each galaxy (either MW or M31) includes contributions from a disk
($\Phi_{disk}$), a bulge ($\Phi_{bulge}$), and a dark matter halo
($\Phi_{NFW}$),
\begin{equation}
\Phi_{tot}(r)=\Phi_{disk}(r)+\Phi_{bulge}(r)+\Phi_{NFW}(r) ,
\end{equation}
where (Xue et al. 2008)
\begin{eqnarray}
\Phi_{disk}(r)=-\frac{GM_{disk}(1-e^{-\frac{r}{b}})}{r}, \\
\Phi_{bulge}(r)=-\frac{GM_{bulge}}{r+c_{0}}, \\
\Phi_{NFW}(r)=-\frac{4\pi G\rho_{s}r_{vir}^3}{c^3r}\ln(1+\frac{cr}{r_{vir}}),
\end{eqnarray}
with
\begin{equation}
\rho_{s}=\frac{\rho_{cr}\Omega_{m}\delta_{th}}{3}\frac{c^{3}}{ln(1+c)-c/(1+c)} .
\end{equation}

We adopt the standard $\Lambda$CDM cosmological parameters $\Omega_m=0.3$,
$H_0=70~\kms\Mpc^{-1}$ (Komatsu et al. 2008), and an overdensity of
$\delta_{th}=340$ (Wechsler et al 2002; Klypin, Zhao, \& Somerville 2002),
with
\begin{equation}
\mvir=\frac{4\pi}{3}\rho_{cr}\Omega_{m}\delta_{th}r_{vir}^{3}\propto
V_{circ}^{3} ,
\label{eqn:mvir}
\end{equation}
and $\rho_{cr}=3H_0^{2}/8\pi G$.

Some of our initial conditions (e.g., \mvir~and \rvir) depend on the
distance of the Sun from the center of the MW galaxy, as will be discussed
in \S \ref{sec:const}, but we consistently adopt
$M_{disk}=4\times10^{10}\msun$, $M_{bulge}=0.8\times10^{10}\msun$, a disk
scale length $b=3.5 ~\kpc$, a disk scale height $c_{0}=0.7 ~\kpc$, and a
halo concentration $c=12$ for the MW; and $M_{disk}=7\times10^{10}\msun$,
$M_{bulge}=1.9\times10^{10}\msun$, $\mvir=1.6\times10^{12}\msun$, $b=5.7~
\kpc$, $c_{0}=1.14 ~\kpc$, $\rvir=300 ~\kpc$, and $c=12$ for M31, as
suggested by Klypin et al. (2002).  Aside from the revised normalization of
\mvir ~and \rvir ~based on the modified value of \vcirc, we have used the
same mass profile as B07.

\section{Constraints on the Orbital History}
\label{sec:const}

Previous calculations of the history of the LMC have adopted the IAU
standard values of $\rsun=8.5~ \kpc$ and $\vcirc=220~\kms$, as derived
by Kerr \& Lynden-Bell (1986; hereafter KLB86).  These values assume
the angular velocity of circular rotation for the Sun to be
$\Theta_{0}/\rsun=25.9~\kms\kpc^{-1}$.  RB04 directly measured this
value\footnote{The supermassive black hole, Sgr A*, is expected to lie
nearly motionless at the dynamical center of the MW since its mass far
exceeds that of the surrounding stars. Hence, the apparent motion of
SgrA* on the sky represents the reflex of the Sun's orbit around the
Galactic center, including both the mean Galactic rotation and the
small peculiar motion of the Sun relative to the local standard of
rest (RB04).}  from the proper motion of Sgr A* to be
$\Theta_{0}/\rsun=29.45\pm.15~\kms\kpc^{-1}$, which using
$\rsun=8\plm.5~\kpc$ (Reid 1993) implies $\vcirc=236\plm15~ \kms$.
Based on the observed orbits of individual stars around Sgr A*, Ghez
et al. (2008) and GGE08 measured the distance of the Sun from the MW
center to be $\rsun=8.4\plm 0.4~ \kpc$, assuming that Sgr A* is at
rest\footnote{Since the surrounding stars are lighter by six orders of
magnitude than Sgr A*, and since there is currently no evidence for a
second (intermediate mass) black hole, the assumption that Sgr
A* is at rest at the dynamical center of the MW appear most
natural.}. These latest measurements are similar to the value inferred
by KLB86, and bring the maximum \vcirc ~(within 1-$\sigma$) up to 265 \kms, which is $\sim
20$\% higher than the 220 \kms~ on which all prior studies were based.
Calculations based solely on KLB86 are therefore at the lower end (2-$\sigma$ below) of
the allowed values for \mvir, \vcirc, and \rvir.  The new values are
listed in Table \ref{tab:MVr}.  The inferred masses and radii assume
the commonly accepted value of $\mvir=1\times10^{12}\msun$ for
$\vcirc=220~ \kms$. The values of $v_{X},~v_{Y},~v_{Z}$ (corresponding
to Galactic Coordinates X, Y, Z) were all calculated using the
standard method of vdM02, changing only the value of \vcirc.

While all of the recalculated masses are consistent with various
recent measurements, the circular velocity of 251~\kms, corresponding
to \mvir=$1.485\times10^{12}\msun$, gives the minimum value of the
virial mass that is consistent with the timing argument, which puts
the total mass of the Local Group (LG) between $3.2\times10^{12}\msun$
and $5.5\times10^{12}\msun$ (Binney \& Tremaine 1987, hereafter BT87;
van der Marel \& Guhathakurta 2008, hereafter vdMG08; Li \& White
2008).  The MW and M31 galaxies dominate the LG and are of comparable
masses; a combined mass of $3$--$4\times10^{12}\msun$ is on the lower
end of the timing argument estimate.  Li \& White (2008), using the
Millennium Simulation and data on Leo I, find the most likely mass for
MW halo to be $M_{MW}=2.34\times10^{12}\msun$ with a lower limit of
$0.8\times10^{12}\msun$, giving a more than adequate range to
acommodate our inferred masses.  Additional mass estimates of the MW
halo are found in Table \ref{tab:MW} and discussed further in \S
\ref{sec:disc}.

\begin{table}
 \caption{Modified Values for \vcirc, \mvir, and \rvir~of the MW}
 \label{tab:MVr}
 \begin{tabular}{@{}ccccc}
  \hline
  \rsun & \vcirc & \mvir  & \rvir & color \\
  (kpc) & (\kms) & ($10^{12}\msun$) & (kpc) & \\
  \hline
  \hline
  7.5 & 220 & 1.000 & 258 & yellow \\
  8.0 & 236 & 1.234 & 277 & green\\
  8.5 & 251 & 1.485 & 295 & blue\\
  9.0 & 265 & 1.748 & 310 & purple\\
  \hline
 \end{tabular}
\end{table}

\begin{figure}
\centering
\includegraphics[width=3in]{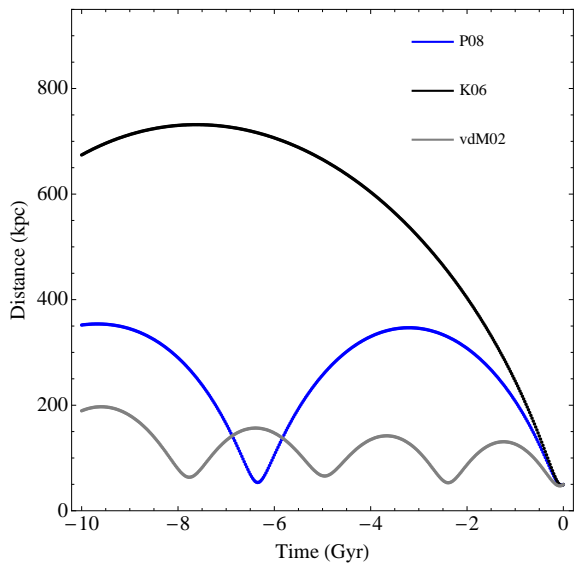}
\caption{Distance of the LMC from the center of the MW galaxy as a
function of time (with zero as the present time) for orbital
velocities published in previous studies (see Table 2), but assuming
$\vcirc=251~\kms$ for the MW.  The blue line is from P08, the gray line is
from vdM02, and the black line is from K06/B07.}
\label{fig:distance2}
\end{figure}

\begin{figure}
\centering
\includegraphics[width=3in]{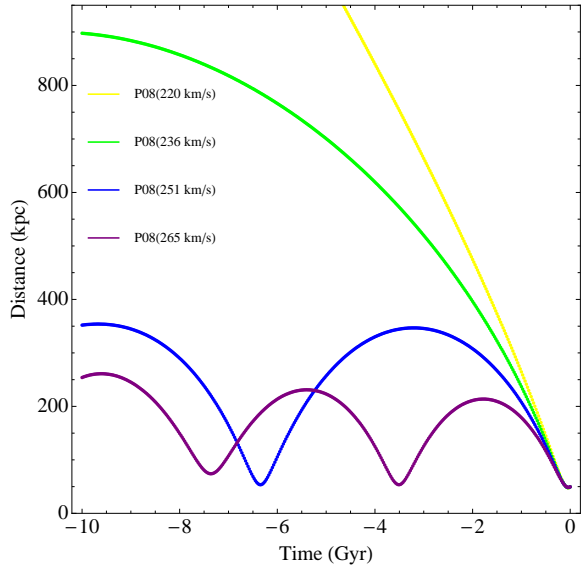}
\caption{Distance of the LMC from the center of the MW galaxy.  The yellow
trajectory (corresponding to $\vcirc=220~\kms$) is gravitationally
unbound to the MW.  The green trajectory (corresponding to
$\vcirc=236~\kms$) is also unbound.  The blue orbit (corresponding to
$\vcirc=251~\kms$) is bound to the MW, with a period of $\sim 6.3$ Gyr
and an apogalacticon of 347 kpc.  The value of \rvir~at this \vcirc~is
295 kpc (see Table \ref{tab:MVr} and equation \ref{eqn:mvir}). The
purple orbit, corresponding to $\vcirc=265~\kms$, is more tightly
bound, with a period of $\sim 3.5$ Gyr and an apogalacticon distance
of 214 kpc.  This distance is well within the virial radius
\rvir$=310$ kpc at this value of \vcirc.}
\label{fig:distance}
\end{figure}

The P08(251~\kms) entry (with this notation denoting the
proper motion from P08 and a circular velocity of 251 \kms) in Table 2
for the LMC velocity of 339 \kms~is significantly lower (by more than
2--$\sigma$) than the K06(220~\kms) value of 378\plm18 \kms, although
it is higher than the vdM02(220~\kms) weighted average of previous
studies, 293\plm39 \kms.  Comparing the corresponding LMC velocities
for $\vcirc=251~\kms$, the K06 and P08 values are much closer (in
agreement with their initial transverse velocities being within one
standard deviation of each other).  P08, K06, and vdM02 with
$\vcirc=251~ \kms$ are all compared in Figure \ref{fig:distance2}.
Figure \ref{fig:distance} compares the P08 proper motion at various
values of \vcirc.

\begin{table}
 \caption{Calculated velocities for the LMC}
 \label{tab:vel}
 \begin{tabular}{@{}lcccc}
  \hline
  Author& \vcirc & $v_{X},v_{Y},v_{Z}$ & $|v|$ & color \\
  ~ & (\kms) & (\kms) & (\kms) &   \\
  \hline
  \hline
  vdM02 & 220 & -56, -220, 186 & 293 &\\
  vdM02 & 251 & -56, -189, 186 & 271 & gray\\
  K06 & 220 & -86, -268, 252 & 377 & \\
  K06 & 251 & -86, -237, 252 & 356 & black\\
  P08 & 220 & -83, -258, 238 & 360 & yellow\\
  P08 & 236 & -83, -243, 238 & 350 & green\\
  P08 & 251 & -83, -227, 238 & 339 & blue\\
  P08 & 265 & -83, -213, 238 & 330 & purple\\
  \hline
 \end{tabular}
\end{table}

\begin{figure}
\centering 
\includegraphics[width=3in]{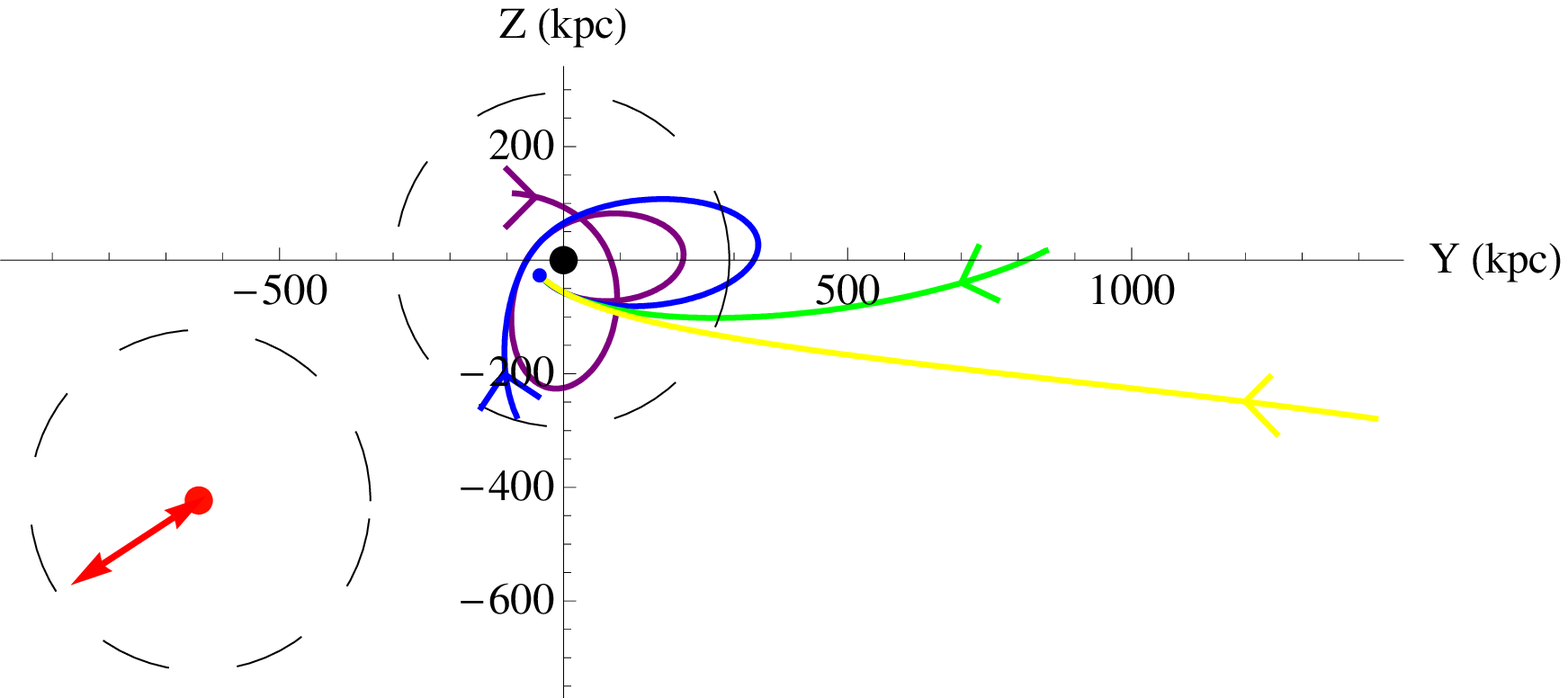}
\caption{Projection of the path of the LMC onto the Y--Z plane
(Galactic coordinates) of the LG over the past 8 Gyr.  The
yellow, green, blue, and purple trajectories correspond to
$\vcirc=220,~236, ~251,$ and $265~\kms$, respectively.  The red line
traces the path of M31 (the large red dot) from and towards the MW.  The
Galactic (MW) center is located at the black dot and the LMC center is
represented by the small blue dot.  The approximate virial radii of the MW and M31
are traced out by the dashed lines.}
\label{fig:2dpathlg}
\includegraphics[width=3in]{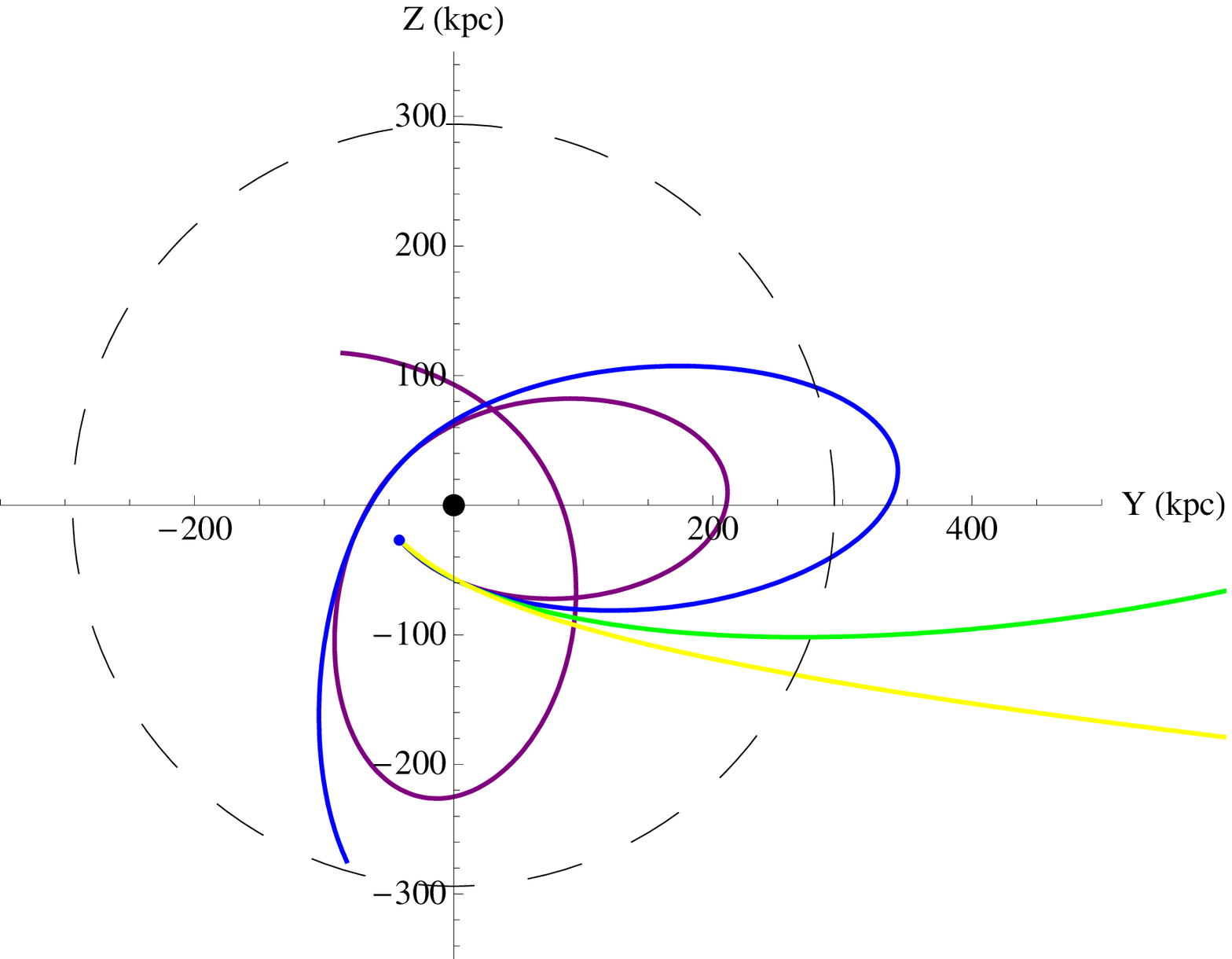}
\caption{A close-up of the MW galactic center in Figure \ref{fig:2dpathlg}.}
\label{fig:2dpath}
\end{figure}

\section{Tidal Effect of M31}
\label{sec:M31}

In our simplified analysis of the dynamics of M31, we ignore its
transverse motion and consider only the perturbative influence of M31
as its radial distance changes relative to the MW (BT87).  While the
transverse component of M31's velocity might be non-negligible (Loeb
et al. 2005), recent analysis suggests that it is lower than the
radial component (vdMG08) and its inclusion would only have a weak
effect on the results reported here.  We also ignore any diffuse
intergalactic mass in between the MW and M31, although future modeling
might take the related uncertainty into consideration (Cox \& Loeb
2007).  To account for the changing separation between the two
galaxies, we used the standard radial dynamics model (BT87).

\begin{figure}
\centering \includegraphics[width=3in]{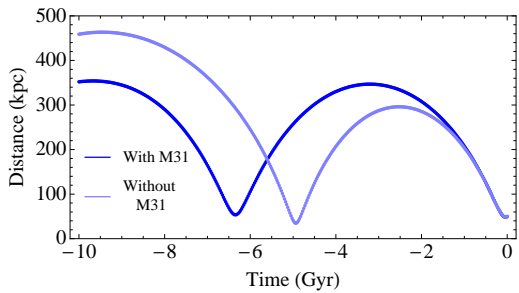} \label{fig:251M31}
\caption{Distance of the LMC from the center of the MW as a function of
time, with (darker line) and without (lighter line) the gravitational
influence of M31, assuming $\vcirc=251~\kms$.}
 \label{fig:M31}
\end{figure}

We find that M31 has a small but non-negligible effect on the trajectory of
the LMC.  Based on the current position of the LMC (see Figures
\ref{fig:2dpathlg} and \ref{fig:2dpath}), the addition of M31 pulls the LMC
away from the Galactic center and towards the LG center of mass.  Figure
\ref{fig:M31} compares the distance of the LMC from the Galactic (MW)
center with (blue line) and without (light blue line) the gravitational
influence of M31, assuming $\vcirc=251~\kms$ for the MW.  Our figures keep
the MW at the center of the coordinate system (corresponding to an
accelerated frame of reference). ~ Without M31, the LMC is more tightly
bound to the MW; in the P08(251\kms) case, the orbital period of the LMC is
5 Gyr, about 1.5 Gyr shorter than if M31 is included.

\section{Discussion}
\label{sec:disc}

We have found that the $\sim 14\pm 6\%$ increase in the MW circular
velocity, relative to the IAU standard of $220~\kms$, allows the LMC to be
gravitationally bound to the MW. Despite its relatively high proper motion
(K06; P08), the orbit of the LMC remains confined within the virial radius
of the MW.

Since the MW and M31 galaxies account for most of the mass in the LG,
the LG itself is not much more spatially extended than the two are
individually, having an estimated zero energy surface at a radius of
$\sim0.9~\Mpc$~(Karachentsev et al. 2008).  If the LMC had followed
the path dictated by the P08(220~\kms) parameters in Table 2, it would
have originated from a distance of $\ga 1.5$ \Mpc~ away from the MW
center (see Figure \ref{fig:2dpathlg}).  In comparison, if we admit
the P08(251~\kms) parameters, then the LMC originated roughly at the
virial radius of the MW halo and well within the boundaries of the LG.
The path suggested by P08(236~\kms) puts the LMC origin outside the LG
but closer than the P08(220~\kms) case.  If the
P08(265~\kms) model is to be believed, the LMC is on its third pass by
the MW center and also originated on the edge of the MW halo.  The
Galactocentric distance on which this result is based
($\rsun=9.0$~kpc) is 1--$\sigma$ above the most recent estimate
(8.4\plm0.5 \kpc; see GGE08), so it is not excluded.

We note that during the long time-scale between perigee passes
($\sim6$ Gyr, see Fig. \ref{fig:distance}), the MW halo mass could
evolve by tens of percent (Diemand, Kuhlen \& Madau 2007). Future
studies might use cosmological simulations to incorporate the
evolution of both the host (MW-like) galaxy and its most massive
(LMC-like) satellite to get a statistical understanding of their
likely past interaction.  Our most likely model has the LMC forming at
roughly the MW virial radius, as expected from a hierarchical
formation of the MW halo (Springel et al. 2008, Figs. 11 \& 12).
Also, all our quoted uncertainty for the LMC orbit stem from the
uncertainty in \rsun (and therefore \vcirc).  Errors in the proper
motion were not taken into account.  The K06 measurements, for
example, are slightly over 1--$\sigma$ off from the P08 measurements
of the LMC proper motion.  The difference between these two cases (as
seen in Figure \ref{fig:distance2}) is drastic -- in one case the LMC
is clearly bound to the MW (P08), and in the other it is not (K06).  A
1--$\sigma$ shift in the other direction, however, would alter the
path of the LMC into an even tighter orbit, similar to the P08(265
\kms) model.

With the value of the distance between the Sun and the Galactic center at
its IAU value (GGE08, KLB86), $\sim$8.5$\plm.5~\kpc$, the total mass of the
MW Galaxy increases by a factor of 1.23--1.75 relative to the values
inferred for the lower distance of 8\plm.5 \kpc ~(Reid 1993).  B07 correctly
rules out a larger mass for the MW but only considers the low mass
($1\times10^{12}\msun$) and the high mass ($2\times10^{12}\msun$) models of
Klypin et al. (2002) which bracket our preferred range.  Xue et al. (2008)
describe the uncertainties of the prior assumption that the galactic \vcirc~
is 220 \kms.  Table \ref{tab:MW} provides the corresponding range of masses
for the MW halo, all calculated from observational data, fit either to a
flat rotation curve or an NFW profile, such as the one we have adopted in
this work.

\begin{table}
 \caption{Milky Way Halo Mass in Recent Studies}
 \label{tab:MW}
 \begin{tabular}{@{}lcc}
  \hline
  Author & Mass & Model \\
  ~ & ($10^{12}\msun$) & ~   \\
  \hline
  \hline
  Wilkinson \& Evans (1999)& $1.9^{+3.6}_{-1.7}$ & FRC$^{1}$ \\
  Sakamoto, Chiba \& Beers (2003) & $2.5^{+0.5}_{-1.0}$ & FRC \\
  Sakamoto, Chiba \& Beers ({\it w/o} Leo I) & $1.8^{+.04}_{-0.7}$ & FRC \\
  Smith et al. (2007) & $1.42^{+1.14}_{-0.54}$ & NFW \\
  Xue et al. (2008) & $0.79\plm0.15$ -  & NFW\\
  ~ & ~~~$1.18\pm0.28$ & \\
  Li \& White (2008)& $2.43$ & N-Body$^{2}$\\
  \hline
 \end{tabular}
$^{1}$FRC denotes a Flat Rotation Curve, as in an isothermal sphere.
These mass calculations have cutoffs of $\sim 50~\kpc$. 
$^{2}$e.g. the Millennium
simulation.  Here the limiting radius is the virial radius,
similar to the NFW calculation.
\end{table}

The $\sim 50 \%$ increase in mass (from $1\times10^{12}~\msun$ to
$1.485\times10^{12}~\msun$) and the $\sim 6\%$ decrease in the
velocity of the LMC (P08(220~\kms) to P08(251~\kms)) have a comparable
influence on making LMC orbit bound to the MW. The combined effect of
these changes is not equivalent to a change in the MW mass, as
considered by B07.  The K06(251~\kms) and P08(220~\kms) lines in
Figures \ref{fig:distance2} and \ref{fig:distance} correspond to
roughly the same LMC velocity (as derived from the IAU standard with
the P08 proper motion) but a different MW mass.  The higher mass of
the MW contributes about half of the overall difference between these
cases.  This parallels the comparison made in B07 between the
K06(220~\kms) and the K06(220~\kms)$+4\sigma$ numbers in their {\it
Fiducial} and {\it High Mass} Models, except that our findings do not
require a $4-\sigma$ deviation from the measured LMC velocity or a
$2-\sigma$ deviation from the mass of the MW in order to make the LMC
orbit bound to the MW.

The previous suggestion (B07) of possibly disqualifying the LMC and
SMC as MW satellites has undesirable implications for the satellites
in the MW halo.  M31 has 18 satellites, 5 of which are gas rich,
whereas the MW, if the LMC and SMC are no longer bound to it, has only
12 bound satellites, 2 of which are gas rich (Karachentsev 2005,
vdMG08).  M31's satellites range in mass from 0.58 to 500
$\times10^{8}\msun$, whereas the remaining MW satellites are in the
range of 0.1--1$\times10^{8}\msun$ (Mateo 1998).  Excluding the LMC
and SMC as satellites of the MW appears unnatural, as there is no
reason for the two comparably sized galaxies to have a large disparity
in the abundance of massive satellites.  Moreover, the chance of
finding massive galaxies like the LMC and the SMC so close to the MW
center requires a special coincidence if they are unbound to the MW,
since they would have spent most of their orbital time far away from
the MW in that case.  Yet, no similar galaxies are known to exist in
the much larger volume between M31 and the MW.  Although these
statistical arguments are not definitive, they support indirectly the
higher updated values of \vcirc (RB04) and \rsun (GGE08) for the MW.

No work on the Magellanic Clouds would be complete without a mention
of the Magellanic Stream (MS).  There is $<2\%$ change in the
trajectory over the length of the MS (100 degrees) between the
P08(220\kms) and the P08(251\kms) models, both of which are well
within the error margins in Figure 8 of B07.  We reiterate the
concerns presented in B07 that neither tidal stripping nor ram
pressure can fully explain the orientation of the Stream.

\noindent
{\bf Acknowledgments.}  We thank Gurtina Besla and Mark Reid for useful
discussions.  This work is supported in part by NASA grant NNX08AL43G,
by FQXi, and by Harvard University and Smithsonian Astrophysical
Observatory funds.


\label{lastpage}
\end{document}